\documentclass[a4paper,12pt]{article}
\pdfoutput=1
\usepackage{graphicx, rotating,amssymb,amsmath,enumerate,latexsym}

\ifx\pdfoutput\undefined
\usepackage[dvips,bookmarks]{hyperref}	
\else
\usepackage{hyperref}	
\fi
\hypersetup{colorlinks,bookmarksopen,bookmarksnumbered,citecolor=rossoc,
linkcolor=verdes,pdfstartview=FitH,urlcolor=rosso}
\def\myurl#1#2{\href{http://#1}{#2}}
\def\hhref#1{\href{http://arxiv.org/abs/#1}{#1}} 

\usepackage{multicol}
\usepackage{color}
\definecolor{rosso}{cmyk}{0,1,1,0.4}
\definecolor{rossos}{cmyk}{0,1,1,0.55}
\definecolor{rossoc}{cmyk}{0,1,1,0.2}
\definecolor{blu}{cmyk}{1,1,0,0.3}
\definecolor{blus}{cmyk}{1,1,0,0.6}
\definecolor{bluc}{cmyk}{1,1,0,0.1}
\definecolor{verde}{cmyk}{0.92,0,0.59,0.25}
\definecolor{verdec}{cmyk}{0.92,0,0.59,0.15}
\definecolor{verdes}{cmyk}{0.92,0,0.59,0.4}

\definecolor{rossoCP3}{cmyk}{0,.88,.77,.40}

\def\FERMI{{\sf Fermi}}
\def\PAMELA{{\sf PAMELA}}
\def\HESS{{\sf H.E.S.S.}}
\def\MAGIC{{\sf MAGIC}}
\def\CTA{{\sf CTA}}
\def\ICECUBE{{\sf Icecube}}

\begin{document}

\setcounter{page}{0}
\thispagestyle{empty}

\begin{flushleft}
\tiny{
CERN-PH-TH/2012-110\hfill 
SACLAY--T12/034\hfill 
IRFU-12-34\hfill
CP$^3$-Origins-2012-10\hfill
DIAS-2012-11\hfill
LAPTH-023/12}
\end{flushleft}

\color{black}
 
\vskip 10pt

\begin{center}

{\bf \LARGE {
Gamma ray constraints\\[0.2cm] on Decaying Dark Matter}}

\color{black}\vspace{0.8cm}

{
{\large\bf Marco Cirelli}$^{\,a,b}$,
{\large\bf Emmanuel Moulin}$^{\,c}$,
{\large\bf  Paolo Panci}$^{\,d}$,\\[0.1cm]
{\large\bf  Pasquale D. Serpico}$^{\,e,a}$,
{\large\bf Aion Viana}$^{\,c}$}

\vspace{0.5cm}

{\footnotesize\noindent
{\it $^a$\href{http://ph-dep-th.web.cern.ch/ph-dep-th/}{CERN Theory Division}, CH-1211 Gen\`eve, Switzerland}\\[0.2cm]
{\it $^b$\href{http://ipht.cea.fr/en/index.php}{Institut de Physique Th\'eorique}, CNRS, URA 2306 \& CEA/Saclay,
	F-91191 Gif-sur-Yvette, France}\\[0.2cm]
{\it $^c$\href{http://irfu.cea.fr/en/}{IRFU/DSM}, CEA Saclay, F-91191 Gif-sur-Yvette, France}\\[0.2cm] 	
{\it $^d$\href{http://cp3-origins.dk/}{CP3-Origins} \& the Danish Institute for Advanced Study { DIAS},\\ 
University of Southern Denmark, Campusvej 55, DK-5230 Odense M, Denmark}\\[0.2cm]
{\it $^e$\href{http://lapth.in2p3.fr/}{LAPTh}, Univ. de Savoie, CNRS, B.P.110, Annecy-le-Vieux F-74941, France}}

\end{center}

\vspace{.1cm}

\begin{abstract}

\vspace{.2cm}

We derive new bounds on decaying Dark Matter from the gamma ray measurements of (i) the isotropic residual (extragalactic) background by \FERMI\ and (ii) the Fornax galaxy cluster by \HESS. 
We find that those from (i) are among the most stringent constraints currently available, for a large range of DM masses and a variety of decay modes, excluding half-lives up to $\sim10^{26}$ to few $ 10^{27}$ seconds. In particular, they rule out the interpretation in terms of decaying DM of the $e^\pm$ spectral features in \PAMELA, \FERMI\ and \HESS, unless very conservative choices are adopted. We also discuss future prospects for \CTA\ bounds from Fornax which, contrary to the present \HESS\ constraints of (ii), may allow for an interesting improvement and may become better than those from the current or future extragalactic \FERMI\ data.

\end{abstract}

\tableofcontents

\newpage

\section{Introduction}
\label{introduction}

The possibility that Dark Matter (DM), which constitutes most of the matter in the Universe, consists of a particle that actually decays on a very long time scale has attracted much attention lately. This is because the decay time scale $\tau_{\rm dec}$ can be taken to be `short' enough that the decay products give signals in current high energy cosmic ray experiments. Namely, if $\tau_{\rm dec} \simeq {\rm few} \cdot 10^{26}$ sec, decaying DM can be invoked to explain the excesses in the fluxes of positrons and electrons measured by \PAMELA, \FERMI\ and \HESS~\cite{proposal decaying DM post PAMELA}. On the other hand, this value of $\tau_{\rm dec}$ is so much longer than the age of the Universe that the slow decay does not make a dent in the overall cosmological DM abundance and does not spoil the agreement with a number of astrophysical and cosmological observations~\cite{astroboundsdecayingDM}~\footnote{See instead~\cite{astrogoodsdecayingDM} for cases in which the decay of DM may actually help cosmology or astrophysics, but for much shorter decay time scales, excluded by the bounds we will present below.}. 

Irrespectively of this recent activity spurred by the charged CR anomalies, decaying DM is an interesting subject by itself. A long time studied case is the one of supersymmetric gravitino DM~\cite{Buchmuller:2007ui}, which is unstable due to R-parity violation. More generically, in several particle physics models, high-scale suppressed operators may naturally mediate the decay of DM (for a miscellaneous list of references see~\cite{Eichler:1989br,Arvanitaki:2008hq,Arina:2009uq}). 

\medskip

From the phenomenological point of view, the main feature of decaying DM with respect to the `more traditional' annihilating DM is that it is less constrained by neutral messenger probes (essentially gamma rays, but also neutrinos) originating from dense DM concentrations such as the galactic center, the galactic halo or nearby galaxies. The reason is simple and well-known: while the signal originating from annihilating DM is proportional to the square of the DM density, for decaying DM the dependence is on the first power; as a consequence, dense DM concentrations shine above the astrophysical backgrounds if annihilation is at play, but remain comparatively dim if DM is decaying. Decaying DM `wins' instead, generally speaking, when large volumes are considered. This is why in the following we will focus on targets as large as galaxy clusters or, essentially, the whole Universe.
 
\medskip

On the observational side, the \FERMI\ and \HESS\  telescopes are making unprecedented progress in the field of  gamma-ray astronomy, producing measurements of many different targets including those of interest for decaying DM. It is therefore a good time to assess the current status of the latter. To this aim, we will compute the predicted signal from decaying DM, for a variety of decay channels (limiting ourselves to 2-body channels, in order to remain as model-independent as possible), and compare it to the gamma-ray measurements, deriving constraints on the decay half-life.
In particular, in this paper, we make use of two distinct probes:
\begin{enumerate}[(I)]
\item \label{dataexg} The isotropic residual gamma ray flux recently measured by \FERMI~\cite{FERMIexg}, which now extends from about 200 MeV up to 580 GeV. The high-energy portion of this measurement is labelled as `preliminary' by the \FERMI\ collaboration, so it should be used with care. However, we note that the data are based on solid premises, since they are obtained with the same procedure already published in~\cite{Abdo:2010nz} with an enlargement of the dataset. We will anyway show later the effect of considering or not the preliminary portion of the data.
\item \label{datafornax} The recent observation in gamma rays of the Fornax galaxy cluster by \HESS~\cite{HESSFornax}.
\end{enumerate}
We choose these probes for a number of reasons, besides the obvious fact that they are among the most recent ones. 
For what concerns the isotropic flux (\ref{dataexg}), it is known since a long time that it represents a powerful testbed for decaying DM (see for instance~\cite{Takayama:2000uz,Overduin:2004sz,Bertone:2007aw}). For what concerns (\ref{datafornax}), we are again motivated by the fact that large virialized objects such as galaxy clusters are a promising target for decaying DM (as briefly mentioned above) and, more in particular, by the fact that indeed stringent constraints have been recently derived using \FERMI\ data~\cite{Zimmer:2011vy,Huang:2011xr}: on the basis of this we explore the constraining power of a complemetary observatory such as \HESS. This is particularly interesting because upcoming \v Cerenkov telescopes offer very promising prospects of improvement in the near future (as opposed to a space based gamma ray observatory such as \FERMI, which will at most increase its statistic by a factor of order 2). Indeed, we will also study the sensitivity of \CTA, the upcoming large \v Cerenkov Telescope Array~\cite{CTA}. 

\medskip

Incremental improvements to the data and the tools relevant for our analysis also comes from other aspects:
\begin{enumerate}[(a)]
\item \label{dataepem} New data on charged Cosmic Rays (CR), which have been recently presented. \FERMI~\cite{FERMIpositrons} has measured the $e^+/(e^++e^-)$ fraction, confirming the notorious rise exposed by \PAMELA\ in 2008~\cite{PAMELApositrons} and extending the measured spectrum to larger energies (about 200 GeV); \FERMI\ has also presented updated measurements of the total flux of $(e^++e^-)$~\cite{Ackermann:2010ij} as well as the measurement of the separated fluxes of $e^+$ and of $e^-$~\cite{FERMIpositrons}; \MAGIC\ has published~\cite{Tridon:2011dk} the measurement of the total flux of $(e^++e^-)$ too; finally, \PAMELA\ has presented results on the flux of pure $e^-$~\cite{Adriani:2011xv}, of $p$~\cite{Adriani:2011cu} and of $\bar p$~\cite{Adriani:2010rc}. 

\item \label{EWcorrections} ElectroWeak (EW) corrections to the DM generated fluxes. 
As discussed in a number of recent works~\cite{EWcorr1}, the emission of the EW gauge bosons ($W^\pm$, $Z$) from the SM particles emerging from the DM decay process can significantly modify the phenomenology. (i) The corrections are particularly relevant for large DM masses (above a TeV); (ii) they can alter significantly the fluxes of charged cosmic rays and gamma rays, both in their spectral shape and in their amplitude, affecting especially the low energies portion~\cite{EWcorr2}. We include EW corrections in all our computations. 

\item \label{improvedprop} An improved propagation scheme for $e^\pm$ in the Galaxy. In semi-analytic treatments of charged CR propagation in the galactic environment, usually employed in DM studies, the energy losses of $e^+$ and $e^-$ are generally considered as space-indepedent and their behaviour with energy $E$ is approximated as $E^2$. However, a space dependence obviously arises from the different conditions (of magnetic field and of ambient light) in  different points of the galaxy and an energy dependence different from $E^2$ occurs in the full Klein-Nishina regime of Inverse Compton scattering. 
We make use of the fluxes of $e^\pm$ as provided in~\cite{PPPC4DMID}, which include these proper space- and energy-dependences. This impacts both the analysis of the $e^\pm$ fits (mildly) and the fluxes of ICS gamma rays from the galactic halo (more significantly).

\end{enumerate}

\medskip

Some recent studies have performed work related to ours~\cite{Chen:2009uq,Ibarra:2009nw,CPS1,Papucci:2009gd,Zaharijas:2010ca,Ke:2011xw,Zimmer:2011vy,Luo:2011bn,Huang:2011xr}. The most closely related, as far as the isotropic gamma ray flux is concerned, is~\cite{CPS1}, in which a subset of us derived constraints on decaying DM with the data and the tools available at that time.  We improve, with respect to that work, by including the updated data in \ref{dataexg} and by the points (\ref{dataepem}), (\ref{EWcorrections}) and (\ref{improvedprop}) discussed above. There are also some other minor analysis differences on which we will comment in the text.
Among other works, ref.~\cite{Huang:2011xr} made a particularly detailed analysis: we consider the more recent data in \ref{dataexg} and \ref{datafornax} above and we include the Inverse Compton Scattering contribution to the gamma ray flux in the galactic component of \ref{dataexg}; the points (\ref{dataepem}), (\ref{EWcorrections}) and (\ref{improvedprop}) are new in our paper; we also consider a few more decay channels. 
For the cases which overlap with this and other studies, we will present a comparison in Sec.~\ref{comparison}.

\medskip

The rest of this paper is organized as follows. 
In Sec.~\ref{charged fits} we update charged CR fits by including (\ref{dataepem}), (\ref{EWcorrections}) and (\ref{improvedprop}). 
In Sec.~\ref{isotropic} we discuss the calculation of the constraints from the isotropic residual $\gamma$-ray flux, while in Sec.~\ref{Fornax} we discuss those from the Fornax cluster.
In Sec.~\ref{results} we present the combined results.
In Sec.~\ref{conclusions} we present our conclusions.


\section{Update of the decaying DM fits to charged CR anomalies}
\label{charged fits}

As mentioned in Sec.~\ref{introduction}, the anomalous \PAMELA, \FERMI\ and \HESS\ data in $e^+$ and $(e^++e^-)$ have been interpreted in terms of DM decay. We recall here briefly the main features of the experimental data and of their DM interpretations, without entering in the details of any specific particle physics model. 

\medskip

We use the following data sets:
\begin{itemize}
\item[$\diamond$] \PAMELA\ positron fraction~\cite{PAMELApositrons}, selecting only points with $E > 20$ GeV, in order both to avoid the low energy region affected by the uncertainty of solar modulation and to have a consistent overlap with the \FERMI\ positron fraction data points (the low energy \PAMELA\ data have very small error bars that would overconstrain the fit).  
\item[$\diamond$] \FERMI\ positron fraction~\cite{FERMIpositrons}.
\item[$\diamond$] \FERMI\ $(e^++e^-)$ total flux~\cite{Ackermann:2010ij}, provided in the low energy (LE) and high energy (HE) samples. 
\item[$\diamond$] \HESS\ $(e^++e^-)$ total flux~\cite{HESSleptons, HESSleptons2}, also provided in a lower energy portion and a higher energy one. 
\item[$\diamond$] \MAGIC\ $(e^++e^-)$ total flux~\cite{Tridon:2011dk}, which however consists of only 6 data points, with error bars larger than those of \FERMI\ and \HESS\ at the same energy and therefore has no effect on the global fit.
\item[$\diamond$] \PAMELA\ $\bar p$ flux~\cite{Adriani:2010rc}. 
\end{itemize}
We perform the fit to these data using the DM generated fluxes as provided in~\cite{PPPC4DMID}, which include the features discussed in points (\ref{EWcorrections}) and (\ref{improvedprop}) above. In looking for the best fitting regions, we scan over the propagation parameters of charged cosmic rays and over the uncertainties on the slope and normalization of the astrophysical electron, positron and antiproton background. 
Note that a simple power law is not expected to necessarily provide a good background model, especially given that the high energy part of the electron spectrum should be dominated by the contribution from local sources. On the other hand, due to the limited number of points available at high-energy and their relatively large statistical error, the background power-law flux is significantly constrained by the lower energy data and the assumption is sufficient in providing a good fit, when combined with the bumpy spectrum from dark matter.
We have assumed a NFW profile for the galactic DM halo (the specific parameters of which are discussed in Sec.~\ref{isotropic}), but other choices would have given almost indistinguishable results. We refer to~\cite{CPS1} and~\cite{CKRS} for more details. 

\medskip

\begin{figure}[!p]
\begin{center}
\hspace{-8mm}
\includegraphics[width=0.50\textwidth]{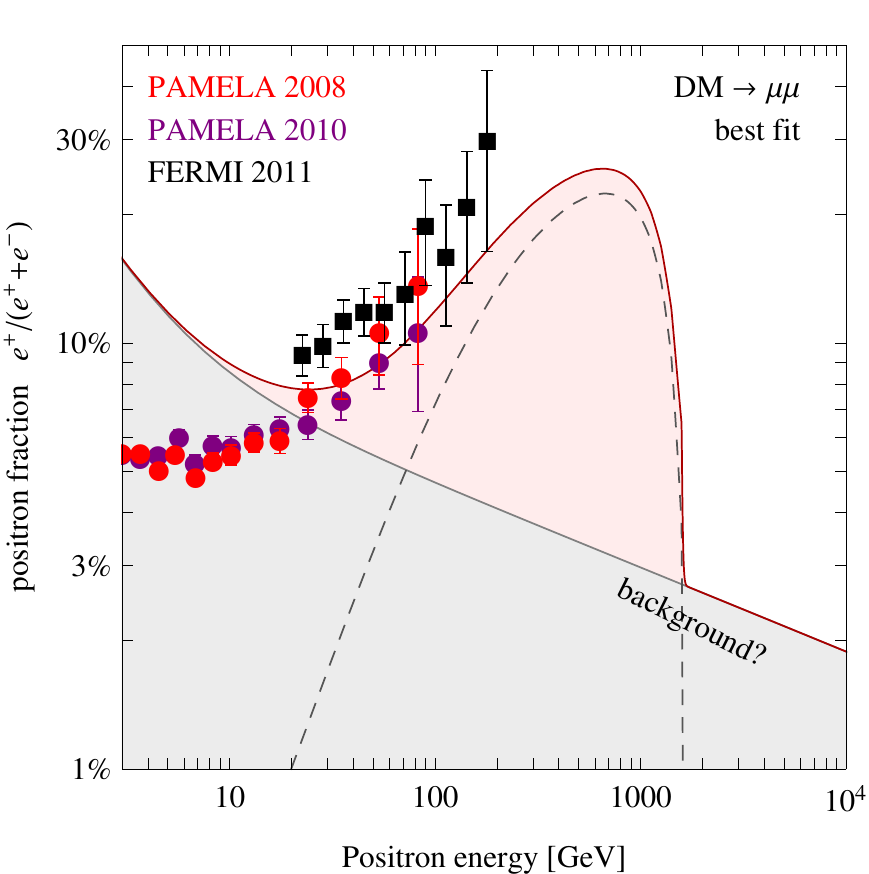}\
\includegraphics[width=0.50\textwidth]{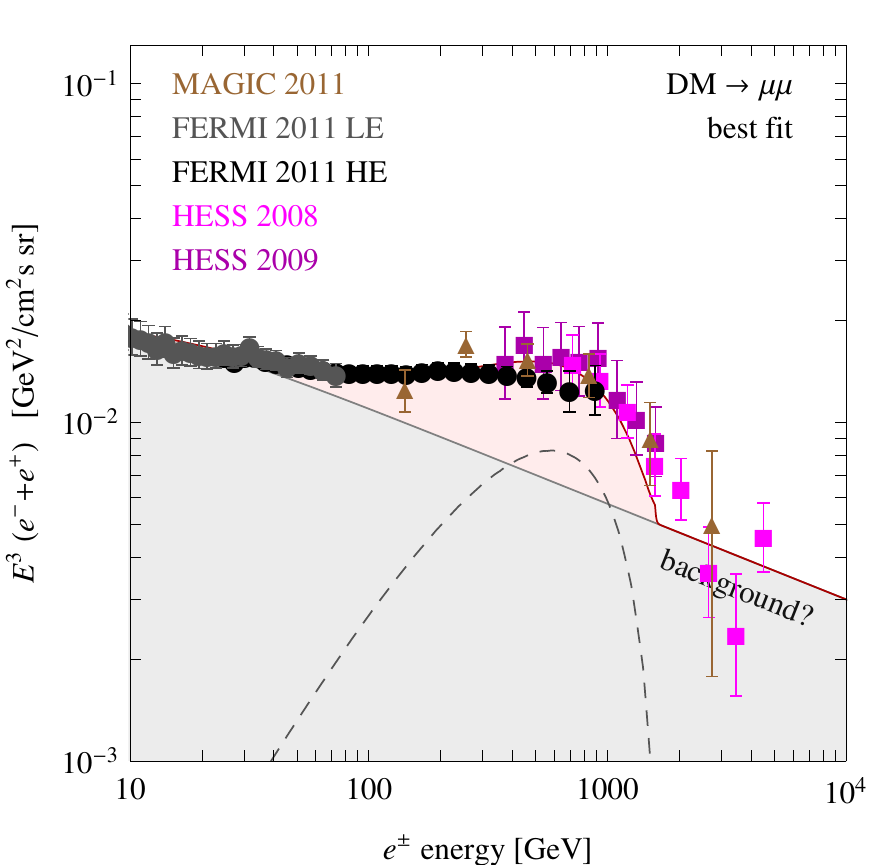}\\
\hspace{-8mm}
\includegraphics[width=0.50\textwidth]{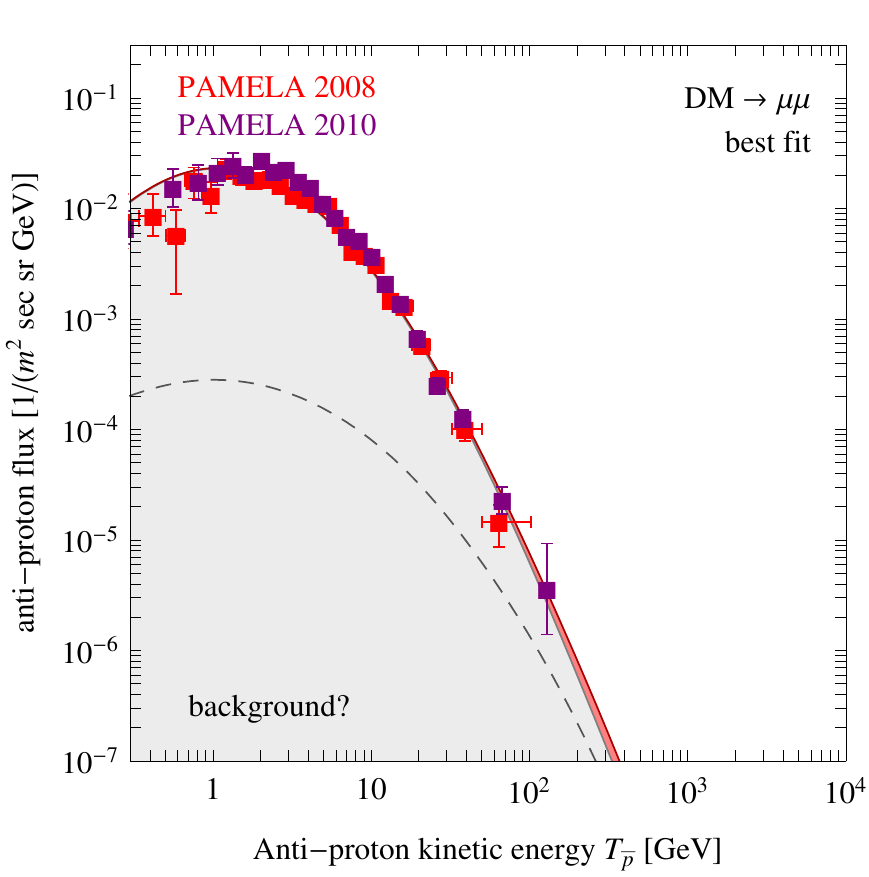}\
\includegraphics[width=0.50\textwidth]{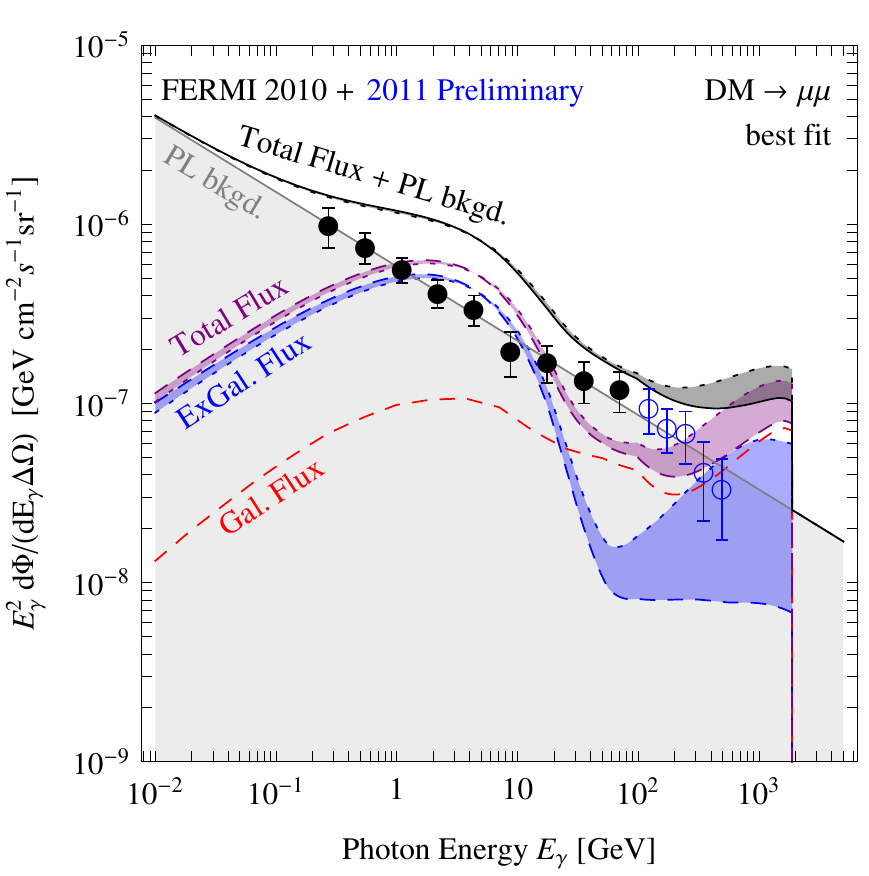}\\
\caption{\em\label{fig:example} An example of the signals in different channels from a decaying DM candidate that fits the charged CR anomalies: we show the positron fraction (upper left), all leptons (upper right), antiproton (lower left) and isotropic gamma ray (lower right) spectra for the best fit candidate DM $\to \mu^+\mu^-$, namely $M_{\rm DM} = 3760$ GeV, $\tau_{\rm dec} = 1.6 \ 10^{26}$ sec (marked by a white cross in fig.s~\ref{fig:exclusionexample} and \ref{fig:exclusion}). The `\,\PAMELA\ 2008' datapoints are reported for completeness, but we use only the more recent `\,\PAMELA\ 2010' in the fits (see text for details). In each panel, the DM contributions are dashed and the astrophysical background is shaded gray. In the lower right panel, the dotted (dashed) lines refer to the flux neglecting (including) extragalactic absorption, see Sec.~\ref{isotropic}.
}
\end{center}
\end{figure}

\begin{figure}[!t]
\begin{center}
\hspace{-8mm}
\includegraphics[width=0.50\textwidth]{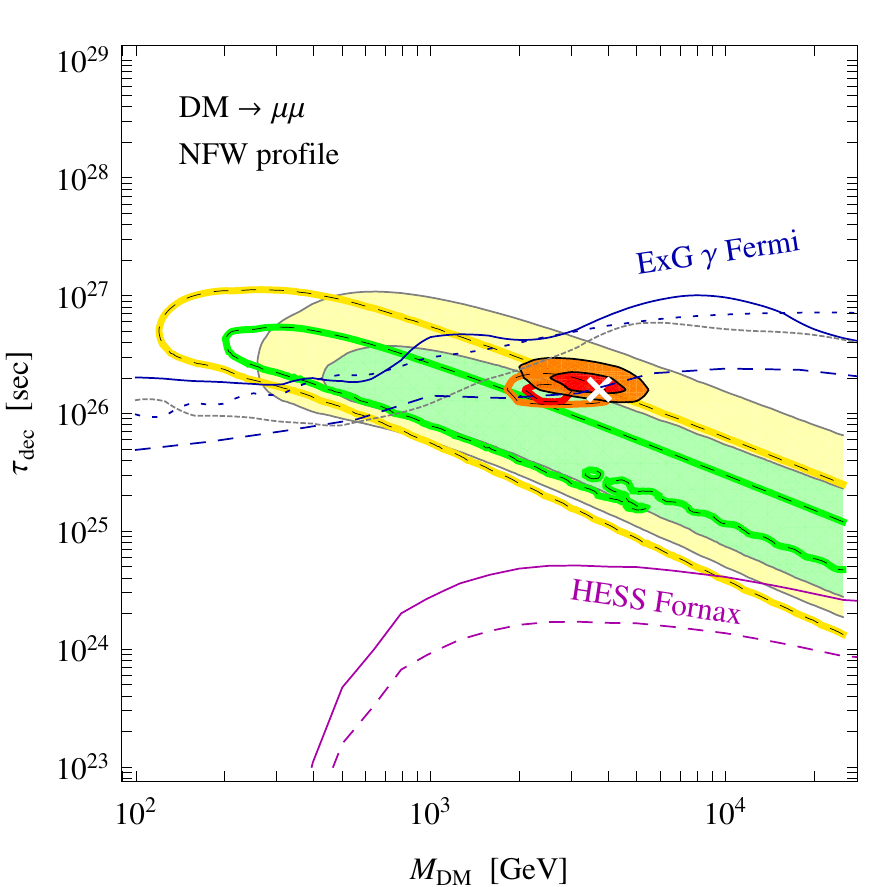}\ 
\includegraphics[width=0.50\textwidth]{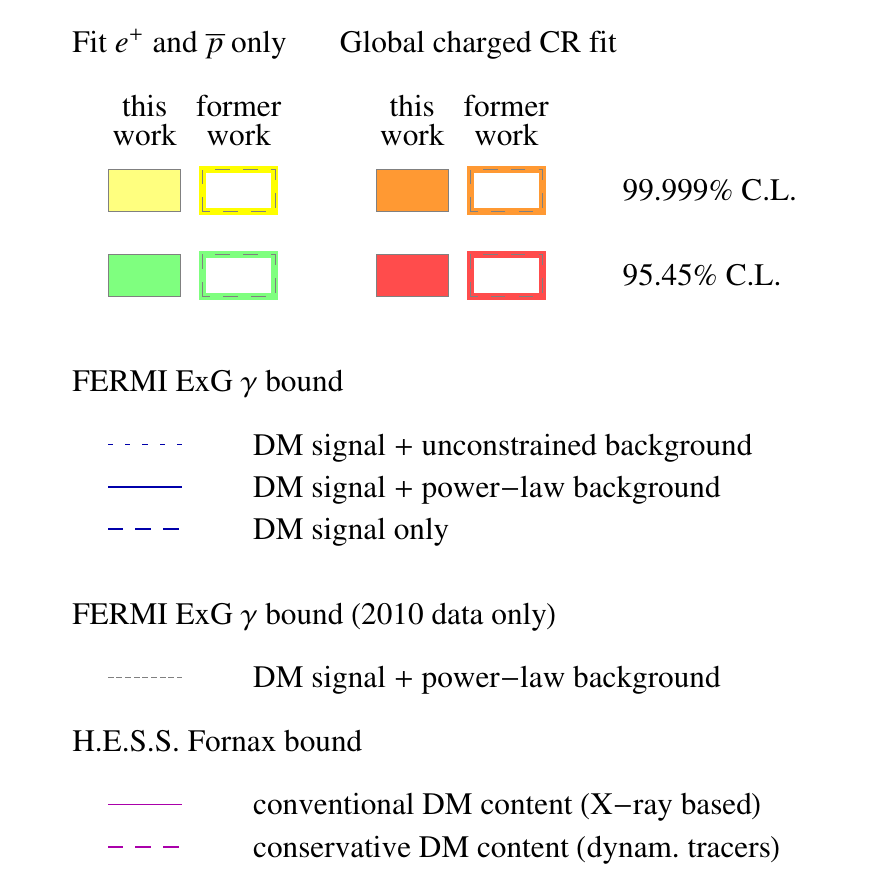}
\caption{\em\label{fig:exclusionexample} 
Illustrative example (for the channel DM$\to \mu^+\mu^-$) of the impact on the fit and constrained regions following from different assumptions and choices, as discussed in the text. In fig.~\ref{fig:exclusion}, on the other hand, we will show only the final regions and our fiducial constraints.
}
\end{center}
\end{figure}

Fig.~\ref{fig:example} presents the spectra results for one specific case: the best fit point for the decay mode DM$\to \mu^+ \mu^-$. 
The allowed regions on the plane $M_{\rm DM}$--$\tau_{\rm dec}$, for the same channel, are instead shown in fig.~\ref{fig:exclusionexample}. We show the 95.45 \%  and 99.999 \% C.L. regions for the fit to positron and antiproton data only (green and yellow bands) and for the whole datasets, i.e. including ($e^++e^-$) (red and orange blobs). 
In fig.~\ref{fig:exclusion}, which we will discuss in more detail later, we show the global allowed regions for all the channels that we consider.

The typical decay time scales that are required for the global fit are of the order of $10^{26}$ to $10^{27}$ seconds. Moreover, as well known~\cite{CKRS}, only `leptophilic' channels allow a global fit: for the quark and gauge boson channels, the few TeV decaying DM needed by ($e^++e^-$) is in conflict with $\bar p$ data.

The impact of the new data and of the improved analysis tools on the identification of the best fit DM properties is overall not big. For instance, in fig.~\ref{fig:exclusionexample} we show the dotted contours of the formerly preferred regions (taken from~\cite{CPS1}), which are not very different. 
However, some features can readily be identified:
\begin{itemize}
\item[-] The fit region for the positron and antiproton signals (yellow and green, see fig.~\ref{fig:exclusionexample}) starts now at larger masses, due to the fact that \FERMI\ $e^+$ data reach higher energies than the former ones from \PAMELA. 
\item[-] An allowed fit region now appears also for the channel DM$\to e^+e^-$ (see fig.~\ref{fig:exclusion}); it was not present in previous analyses such as~\cite{CPS1}. This is due to a number of concurring reasons, among which the fact that the inclusion of the EW corrections (\ref{EWcorrections}) and the refined propagation (\ref{improvedprop}) smoothens out the $e^++e^-$ spectrum and allow a decent fit to the $\sim$ 1 TeV hump, the fact that we now remove from the fit the low energy $e^+$ \PAMELA\ datapoints and, also, that we are now employing for \FERMI\ data the full error bars (slightly larger than before).
\item[-] EW corrections imply the presence of a non-zero $\bar p$ flux even for leptonic decay channels (see e.g. the third panel of fig.~\ref{fig:example}). However, the yield is not enough to appreciably affect the fit regions. 
\end{itemize}
A comment on the selection of the datasets and their compatibility is also in order. We use the full set of data listed at the beginning of this section, but we also checked that dropping the data by \PAMELA\ on the positron fraction, i.e. using only data from \FERMI\ and \HESS\ (on the positron fraction and the pure $e^+$ and $e^-$ fluxes), does not significantly change the fit regions. Similarly, the regions are not significantly modified if we adopt for the positron astrophysical background the recent determination in~\cite{Lavalle:2010sf} (instead of the one conventionally used, from~\cite{Moskalenko:1997gh,Baltz:1998xv}). This is expected, since the predictions in~\cite{Lavalle:2010sf} and in~\cite{Moskalenko:1997gh,Baltz:1998xv} are very similar in the range of energies in which we are interested.


\section{Isotropic gamma ray flux}
\label{isotropic}

The measurements in (\ref{dataexg}) by the \FERMI\ satellite correspond to the (maximal) residual, isotropic gamma-ray flux present in their data. Its origin can be in a variety of different phenomena, both in the form of unresolved sources and in the form of truly diffuse processes (see~\cite{FERMIexg} and reference therein). 

DM decays can also contribute to this isotropic flux, with two terms: 1) an extragalactic cosmological flux, due to the decays at all past redshifts; 2) the residual emission from the DM halo of our Galaxy.
The former is of course truly isotropic, at least as long as one neglects possible nearby DM overdensities. The latter is not, but its minimum constitutes an irreducible contribution to the isotropic flux. 
In formul\ae, the predicted differential DM flux that we compare with \FERMI\ isotropic diffuse $\gamma$-ray data is therefore given by 
\begin{equation}  
\frac{d \Phi_{\rm isotropic}}{d E_\gamma} = \frac{d \Phi_{\rm ExGal}}{d E_\gamma} + 
4 \pi \left. \frac{d \Phi_{\rm Gal}}{d E_\gamma\, d \Omega}\right|_{\rm minimum}
\label{eq:main}
\end{equation}  
For typical DM decay channels and for any DM halo profile, we find that the two contributions are of comparable amplitude, as it can be seen e.g. in the lower right panel of fig.~\ref{fig:example}. We now move to discuss the calculations of the two contributions separately, reproducing and updating the discussion of~\cite{CPS1}.

\medskip

The extragalactic flux is given, in terms of the Earth-measured photon energy $E_\gamma$, by 
\begin{equation}
\frac{d \Phi_{\rm ExGal}}{d E_\gamma} = \Gamma_{\rm dec}\, \frac{\Omega_{\rm DM}\,\rho_{c,0}}{M_{\rm DM}} 
\,\int_0^\infty d z\,\frac{e^{-\tau(E_\gamma(z),z)}}{H(z)} \frac{d N}{d E_\gamma}(E_\gamma(z),z)\,,
\label{smoothedmap}
\end{equation}
where $\Gamma_{\rm dec}= \tau_{\rm dec}^{-1}$ is the decay rate.
Here the Hubble function $H(z) = H_0\sqrt{\Omega_M(1+z)^3+\Omega_\Lambda}$, where $H_0$ is the present Hubble expansion rate. $\Omega_{\rm DM}$, $\Omega_M$ and  $\Omega_\Lambda$ are respectively the dark matter, matter and cosmological constant energy density in units of the critical density, $\rho_{c,0}$. 
The gamma ray spectrum $d N/d E_\gamma$, at any redshift $z$, is the sum of two components: (i) the high energy contribution due to the prompt $\gamma$-ray emission from DM decays and (ii) the lower energy contribution due to Inverse Compton Scatterings (ICS) on CMB photons of the $e^+$ and $e^-$ from those same decays. 
Using eq.~(\ref{smoothedmap}), the extragalactic flux can therefore be computed in terms of known quantities for any specified DM mass $M_{\rm DM}$ and decay channel. 
In particular, since the flux essentially consists of the integral over $\rho$, at the first power, the result is not affected by the formation of DM halos, the history and the properties of which are highly uncertain. This is in contrast to the case of annihilating DM, where instead these issues affect the predictions by orders of magnitude (see e.g.~\cite{Abdo:2010dk,Serpico:2011in}).
We take the resulting fluxes from~\cite{PPPC4DMID}, but we improve the treatment of the effect of the finite optical depth of the Universe, as we now move to discuss.

The factor $e^{-\tau(E_\gamma,z)}$ in eq.~(\ref{smoothedmap}) accounts for the absorption of high energy gamma rays due to scattering with the extragalactic UV background light: essentially, a high energy photon from DM hits a lower energy UV photon and produces an energetic electron (or positron) which, in turn, makes Inverse Compton Scattering on the CMB and therefore reinjects a continuum of lower energy $\gamma$-rays. The process amount to a subtraction of $\gamma$-ray flux from the higher energies and a redistribution towards the lower part of the spectrum. As an element of novelty in this analysis, we take this process fully into account, using the energy and redshift dependent optical depth of the Universe (denoted $\tau(E_\gamma, z)$) of~\cite{PPPC4DMID}. The effect is sizable at energies $E_\gamma \gtrsim 100\,$GeV: it can reduce the high energy flux by about one order of magnitude.
Indeed, in fig.~\ref{fig:example}, lower right panel, we show with dotted lines the fluxes computed neglecting absorption and with dashed lines the effect of including it. Summing over the galactic component (which is of course not affected by absorption) reduces the impact of the effect, which can however remain sizable. This will affect the constraints derived below, especially for channels which feature a large prompt contribution.

\medskip

The flux from the galactic halo, coming from a generic direction $d \Omega$, is given by the well known expression 
\begin{equation}
\frac{d \Phi_{\rm Gal}}{d E_\gamma \ d \Omega} = \frac{1}{4\pi} \frac{\Gamma_{\rm dec}}{M_{\rm DM}} \int_{\rm{los}} d s \, \rho_{\rm halo}[r(s,\psi)] \, \frac{d N}{d E_\gamma}\,,
\label{fluxdec}
\end{equation}
i.e. as the integral of the decaying DM density piling up along the line of sight individuated by the direction $d \Omega$. Here $\rho_{\rm halo}$ is the DM distribution in the Milky Way, for which we will always take a standard Navarro-Frenk-White~\cite{Navarro:1995iw} profile 
\begin{equation}
 \rho_{\rm NFW}(r)=\rho_{s}\frac{r_{s}}{r}\left(1+\frac{r}{r_{s}}\right)^{-2},
   \label{eq:NFW}
\end{equation}
with parameters $r_{s} = 24.42$ kpc and $\rho_{s} = 0.184$ GeV/cm$^3$~\cite{PPPC4DMID}. In principle one could consider other choices of profiles and parameters, but, for the case of decaying DM and considering that we will not be interested in the regions of the Galactic Center (GC) where profiles differ most (quite the opposite: we will focus on the anti-GC), these choices make a negligible difference.
The coordinate $r$, centered on the GC, reads $r(s,\psi)=(r_\odot^2+s^2-2\,r_\odot\,s\cos\psi)^{1/2}$, where $r_\odot = 8.33$ kpc is the most likely distance of the Sun to the GC and $\psi$ is the angle between the direction of observation in the sky and the GC.

The spectrum $d N/d E_\gamma$ consists again of two components: the prompt one and the ICS one. We compute both using the tools in~\cite{PPPC4DMID}. In particular, for the ICS flux we use the generalized halo functions for the IC radiative process provided there, which take into account the full energy losses discussed in (\ref{improvedprop}).
This is another refinement with respect to~\cite{CPS1}, which had computed the $e^\pm$ energy losses towards the anti-GC including CMB only: it was thus missing synchrotron losses (which can account for up to roughly a third of the total losses, in the range 8.33 kpc $\lesssim r \lesssim$ 15 kpc) and ICS losses on local ambient light other than the CMB (which are instead less important). The treatment that we now adopt represents the most refined one we can afford, at the current state-of-the-art of these semi-analytical computations.

As indicated in eq.~(\ref{eq:main}) and in the discussion above it, we need to determine the minimum of the flux in eq.~(\ref{fluxdec}). 
For the prompt contribution, the minimum is obviously located where the line-of-sight density of DM is the lowest, just because the $\gamma$-ray emission density traces the DM density by definition. This coincides with the direction of the anti Galactic Center (anti-GC). 
For what concerns the ICS contribution, on the other hand, the situation is more complicated. In this case the source of $\gamma$-rays is the population of DM-originated $e^\pm$, and therefore the minimum depends on what is assumed for the propagation and final distribution of the latter ones in the galactic halo. For instance, assuming a thin diffusive halo with $L = 1$ kpc (see e.g.~\cite{PPPC4DMID} for a discussion and an overview of standard values) implies that the layer of ICS emitting $e^\pm$ is particularly shallow in the zenith (and nadir) direction and therefore the ICS $\gamma$-ray flux is minimal along lines of sight pointing directly above (and below) the solar system. On the other hand, for a thick $L = 15$ kpc diffusive halo, the ICS emission from the directions orthogonal to the galactic plane is greatly enhanced and the minimum is located again at the anti-GC. In view of this complicated morphology, we choose to locate the minimum always at the anti-GC:
\begin{equation}  
 \left. \frac{d \Phi_{\rm Gal}}{d \epsilon\, d \Omega}\right|_{\rm minimum} \longrightarrow  \left. \frac{d \Phi_{\rm Gal}}{d \epsilon\, d \Omega}\right|_{\rm anti-GC}
\label{eq:minimum}
\end{equation}  
This is supported, along the lines of the discussion above, by a number of considerations, among which: (i) thin diffusive halos are anyhow disfavored by other arguments (see e.g.~\cite{Bringmann:2011py}); (ii) a number of other uncertain astrophysical variables would enter in a detailed determination of a true minimum: for instance if the {\em radial} size of the diffusive halo is smaller than usually assumed ($\sim$20 kpc), then the ICS flux at the anti-GC is reduced; (iii) this discussion only affects the galactic ICS contribution (not the galactic prompt and the extragalactic contributions), thus limiting the impact on the derived total flux and having an importance (for the bounds we will derive below) only for those configurations in which the ICS saturates the constraints.

\subsection{Deriving the constraints}

We thus compute $d \Phi_{\rm isotropic}/d E_\gamma$ as discussed above and we compare it with the \FERMI\ data of (\ref{dataexg}). We show an example of such a comparison in Fig.~\ref{fig:example}, lower right panel. It is pretty clear (even by inspection, in the case of this example) that the DM signal does not agree in shape with the data, which are instead well fit by a simple power law~\cite{FERMIexg}. For this reason, we are driven to derive constraints on the maximum DM signal, and therefore the minimum $\tau_{\rm dec}$, admitted by the data. There are however several possible ways to compute such constraints. We discuss a few of them in the following and we illustrate them in the specific example of fig.~\ref{fig:exclusionexample}.
\begin{itemize}
\item[$\circ$] {\em DM signal only.} The most conservative option consists in demanding that the DM signal alone (eq.~(\ref{eq:main})) does not exceed any single one of the \FERMI\ data points by more than a chosen significance, which we set at 3$\sigma$. 
This is overly conservative for a number of reasons. First of all, since it supposes a vanishingly small astrophysical background. A simple inspection of the DM prediction proves that this assumption is physically untenable;  the bumpy shape of the signal is so different from the featurless observations that the 
(bulk of the) latter must be explained by  astrophysical processes, leaving only a subleading role for signals from DM. Second, given the smooth nature of the DM signal and of the data, it is clear that an excess of 3$\sigma$ in one point is often accompanyed by similar excesses in neighboring points, so that the `global significance' of the exclusion is actually higher. 
For instance, the best-fit model example in fig.~\ref{fig:example} is only barely excluded by such a procedure, despite the clear tension with several data points. 

\item[$\circ$]  {\em DM signal + power-law background.} A more realistic option consists in assuming that the astrophysical background consists of a power-law and demand that the sum of astrophysical background and DM signal does not exceed a chosen global significance. 
Note that one expects several potential sources to contribute
to this flux, like unresolved blazars, star-forming galaxies or electromagnetic cascades from ultra-high energy cosmic ray losses. In general, a combination of them is not
expected to produce (a priori) an exact power-law, which may be inaccurate also for some of the different contributions taken alone. Nonetheless, within the current precision we find that the data of $E_\gamma^2 d\Phi/dE_\gamma \, d\Omega$ alone are very well described by a power law with index -0.41 (very similar to the one found in resolved blazars) 
and normalization 1.02 $\, 10^{-5}$ GeV/(cm$^2$ sec sr).
Next, for a given $M_{\rm DM}$ we add the DM signal, whose normalization is controlled by $\Gamma_{\rm dec}$. We let the normalization of the power law background vary within a factor of 2 from the central value specified above and its index within 0 and -1 (this choice is astrophysically plausible, although varying the parameters in a broader range would not change the results shown). We then compute the $\chi^2$ to the data, marginalizing over these parameters. We compute in this way 95\% C.L. limits on $\Gamma_{\rm dec}$.  

\item[$\circ$] {\em DM signal + unconstrained background.} A variation of the procedure described in the previous point consists in avoiding to commit on the functional form of the background and assume that it consists of an arbitrary function which describes well the data (namely: it is consistent with each data point within a small significance, e.g. 1$\sigma$). We then demand that, adding the signal, the $\Delta\chi^2$ does not exceed a given value, which in this case we fix very conservatively at 25. 
For the practical case at hand, however, this procedure and the previous one yield similar results, as one can see on the example in fig.~\ref{fig:exclusionexample}. This is not surprising, since the isotropic residual datapoints are indeed very well described by a power law.  
\end{itemize}
The procedure {\em `DM signal + power-law background'} is the one that we will adopt to obtain the fiducial constraints shown in our results (see fig.~\ref{fig:exclusion}). This procedure is `fiducial' also in the sense that it matches the analysis we do for charged CR anomalies (see Sec.~\ref{charged fits}) and therefore fit regions and constraints are essentially consistent with each other in fig.~\ref{fig:exclusion}.

\medskip

Before moving on, we discuss how the constraints would be modified if we removed from the analysis the datapoints above 100 GeV, that are labelled as `preliminary' by the \FERMI\ collaboration, i.e. if we limited the data to those published during 2010 in~\cite{Abdo:2010nz} only. In fig.~\ref{fig:exclusionexample} the dotted gray line corresponds to how the fiducial constraints would be modified. We see that, as expected, the limits become looser by a factor of a few.  


\section{Fornax cluster gamma ray flux}
\label{Fornax}

Galaxy clusters are the largest gravitationally bound structures in the universe,  80\% of their total mass being in the form of Dark Matter. 
Although they are located at much larger distances than other popular targets such as dwarf satellites of the Milky Way, they turn out to be attractive environments to search for DM due to their predicted high DM luminosity. 
In practice, which cluster is the most promising depends on a number of factors, and most notably the astrophysical background. Indeed, standard astrophysical gamma-ray emission is expected, both from high energy 
cosmic rays interacting in the intra cluster medium (the dominant contribution) or from electrons in the ambient radiation field (a subdominant contribution, see~\cite{Blasi:2007pm} for a review). Despite these predictions, no gamma-ray from these processes has been detected so far~\cite{Aharonian:2008uq,Aharonian:2009bc,2010ApJ...710..634A,2010JCAP...05..025A,2010ApJ...717L..71A,HESSFornax}. 
Radio galaxies lying at the gravity center of clusters such as M87 in Virgo and NGC1275 in Perseus are also $\gamma$-ray emitters: these have actually been directly observed. 

\smallskip

A few galaxy clusters have been observed by \HESS: the most attractive of them for DM searches are Virgo, Coma and Fornax. 
Although Virgo is closeby, the high energy gamma-ray emission from M87 prevents searches in the inner region. Coma is among the most massive galaxy clusters but it has been shown that it is not a privileged environment because of a relatively high CR-induced gamma-ray emission~\cite{Pinzke:2009cp,2011PhRvD..84l3509P}. Fornax is located at 19 Mpc~\cite{Reiprich:2001zv} and shows favorably-low expected astrophysical 
background~(see, for instance, \cite{2009PhRvD..80b3005J,Pinzke:2009cp,2011PhRvD..84l3509P}). Given its location
near the tropic of Capricorn, the \HESS\ instrument is best-suited to observe Fornax with respect to other currently-operating Imaging Atmospheric \v Cerenkov Telescopes (IACTs). 

\smallskip

The predicted DM $\gamma$-ray flux from Fornax can be easily obtained by integrating Eq.(\ref{fluxdec}) over the observational solid angle $\Delta\Omega$, and of course replacing $\rho_{\rm halo}$ with $\rho_{\rm Fornax}$, the DM distribution in the cluster. 

There are a few different methods to determine $\rho_{\rm Fornax}$ (and in general the DM content of galaxy clusters). 
A widely used approach is based on X-ray measurements on the gravitationally bound hot intracluster gas.
Assuming a NFW profile~\cite{Navarro:1995iw}, the latter is entirely defined from the virial mass extracted form the HIFLUGCS catalog catalog~\cite{Reiprich:2001zv}  and 
the virial mass and the concentration relationship found in Ref.~\cite{Buote:2006kx}, in $\Lambda$CDM cosmology.
Other choices for the profile are possible, e.g. the cored Burkert profile~\cite{Burkert:1995yz}, which may be particularly motivated by the fact that baryon physics in the inner part of clusters may significantly alter the predictions of $\Lambda$CDM simulations: several processes invoking baryons in galaxy formation such as dynamical friction, AGN feedback or gas outflows, may flatten the DM cusps in 
DM cores~\cite{2001ApJ...560..636E,Martizzi:2011aa}. 
For the case of Fornax, various dark matter halo models have been considered in~\cite{HESSFornax} (and references therein). 
However, as opposed to the annihilation case, the gamma-ray flux from decaying DM is less dependent 
on the DM distribution inside the target due to the simple linear dependence of the astrophysical factor with DM density.  
Varying over the profiles discussed in~\cite{HESSFornax}, the difference in the flux factor 
is less than a factor of 3 for a given opening integration angle. 
Another approach to determine the DM content of clusters consists in using dynamical tracers (see again~\cite{HESSFornax} and references therein).  

In the following, we take as `fiducial' the X-ray based determination of the DM profile, hereafter referred to as the RB02 profile~\cite{HESSFornax}, which also well agrees with tracer dynamics at large distances. In this case, the DM distribution is inferred from a X-ray concentration-virial mass relationship\,\footnote{The virial galaxy cluster mass $M_{\Delta}$ is usually defined as the enclosed mass within a radius where the density reaches $\Delta\times\rho_{\rm c}$, with $\rho_{\rm c}$ the critical density of the universe and $\Delta$  taking values between 100 and 500. Although $M_{\rm 500}$ can be used to derive specific galaxy cluster properties, the X-ray concentration-mass relation is particularly well verified for $\Delta$ = 200~\cite{Voit:2004ah}.} and the uncertainty on the mass determination is about 10\%~\cite{Buote:2006kx}. 
As an alternative choice, we also consider the DW01 profile of~\cite{HESSFornax}, which is fully based on the dynamical tracers method. This generates a predicted DM $\gamma$-ray flux $\sim$3 times smaller than RB02.

Next, we discuss the choice of $\Delta \Omega$. In contrast to annihilating dark matter for which most often the smallest opening angle provide the most sensitive searches, at least for cuspy profiles, decaying dark matter 
searches require optimization of the opening angle to guarantee the highest signal-to-noise ratio.  As it is straightforward from Eq.(\ref{fluxdec}), the luminosity scales  with the size of the solid
integration angle. On the other hand, background is increasing as well. We find that the optimization of the signal-to-noise ratio versus the opening integration angle for the dark matter halo profile RB02 implies the best integration region to be 0.5$^{\circ}$, which corresponds to a solid angle of $\rm \Delta\Omega = 2.4\times10^{-4}$ sr.          

\medskip

We also mention that an additional component to the decaying Dark Matter spectrum needs in principle to be considered, particularly for leptonic final states: that of Inverse Compton emission (indeed, as opposed to dwarf galaxies, in clusters the electrons lose energy primarily through ICS on the ambient radiation field and produce additional gamma rays in the final state). We include it in the computations for the DM $\to \mu^+\mu^-$ channel, but, due to the energy working range of \HESS, this component becomes important only for very large DM masses (above $\sim\,$30 TeV) and its effect is therefore not visible in our results.

\subsection{Deriving the constraints}

In order to extract exclusion limits on the Dark Matter lifetime from gamma-ray astronomical observation with IACTs, the background needs to be determined to constrain the decaying DM luminosity. The background is calculated in a region referred to as the OFF region, and the signal region as the ON region.
Both ON and OFF regions depend on the observation mode and are specific to the IACT instrument.  As for the background level, it is taken simultaneously, {\it i.e.} in the same data-taking observing conditions, to the signal events in order to allow for the most accurate estimate. 

In the `standard' ON-OFF method, one has to take into account the background flux in the astrophysical factor and subsequently in the upper limit calculation. For the ON region of 0.5$^{\circ}$ and considering a OFF region in an annulus around the ON region with an inner and outer radii of 0.5$^{\circ}$ and 1.0$^{\circ}$ respectively, this reduces the decaying DM flux in the ON region of about 40\%. An `improved' background estimate procedure called the {\it template} method allows to avoid this reduction. In such method the background events are determined in the ON region but from selecting hadron-like events (see~\cite{HESSFornax} for details on the analysis procedure). We indeed make use of it in this work.

\medskip

\HESS\ has observed Fornax for a total of 14.5 hours at low zenith angle to allow for best sensitivity to low dark matter masses.


\section{Results and discussion}
\label{results}

\begin{figure}[!p]
\begin{center}
\hspace{-8mm}
\includegraphics[width=0.45\textwidth]{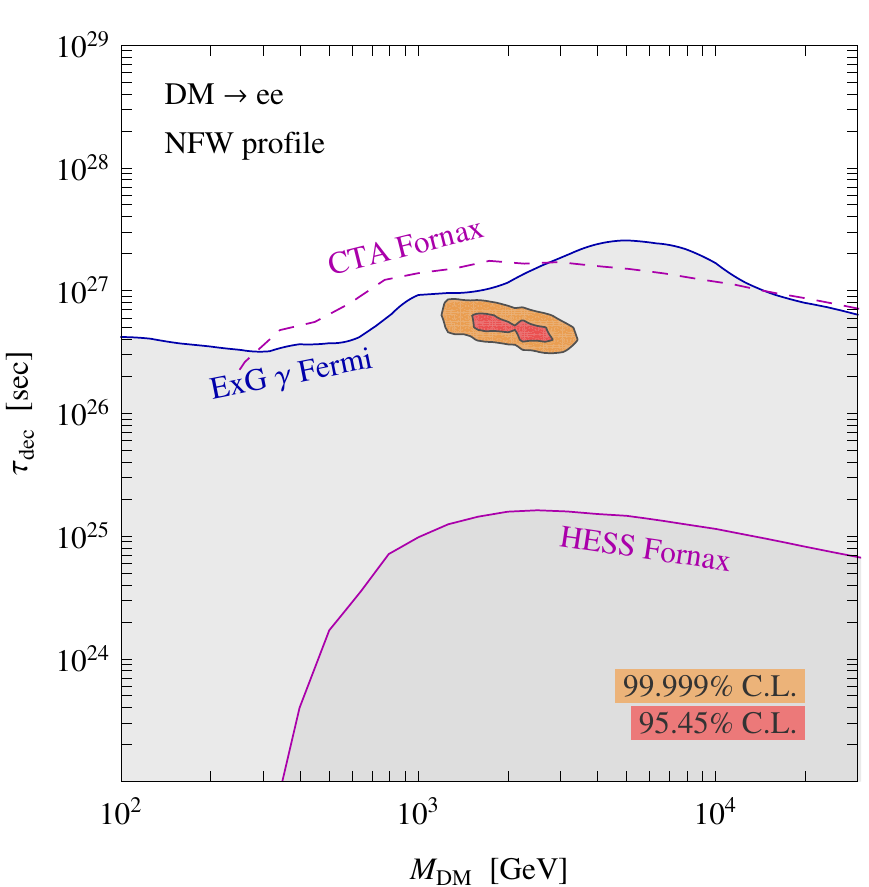}\
\includegraphics[width=0.45\textwidth]{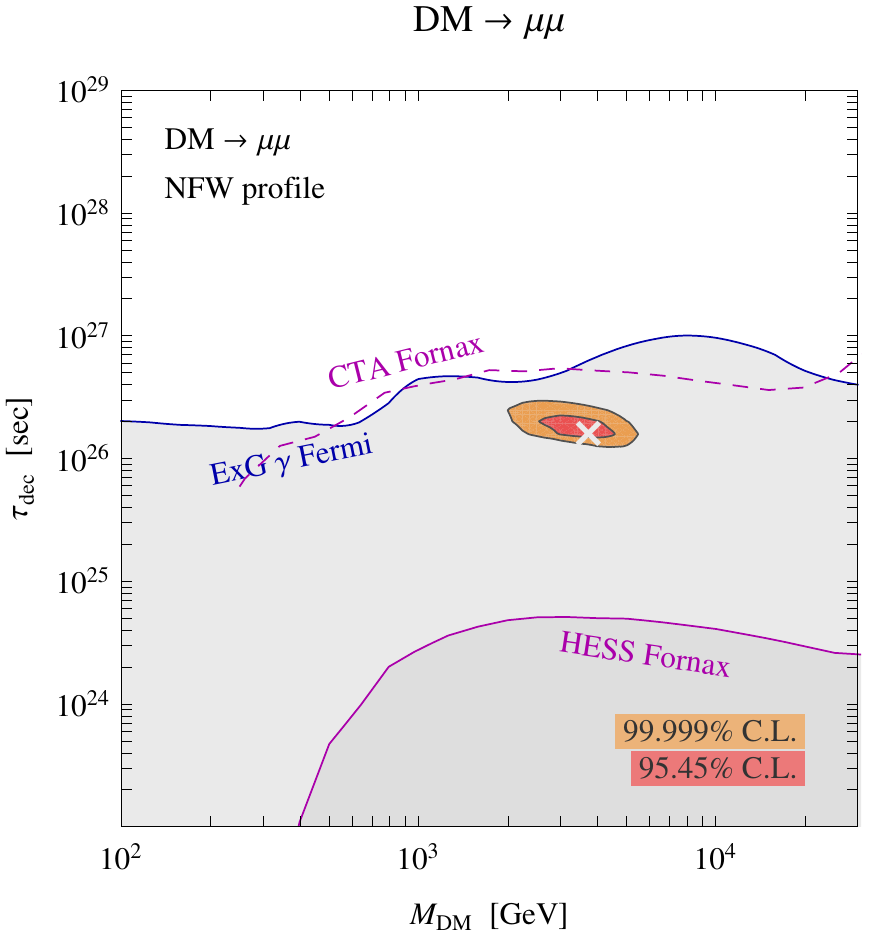}\\[1mm]
\hspace{-8mm}
\includegraphics[width=0.45\textwidth]{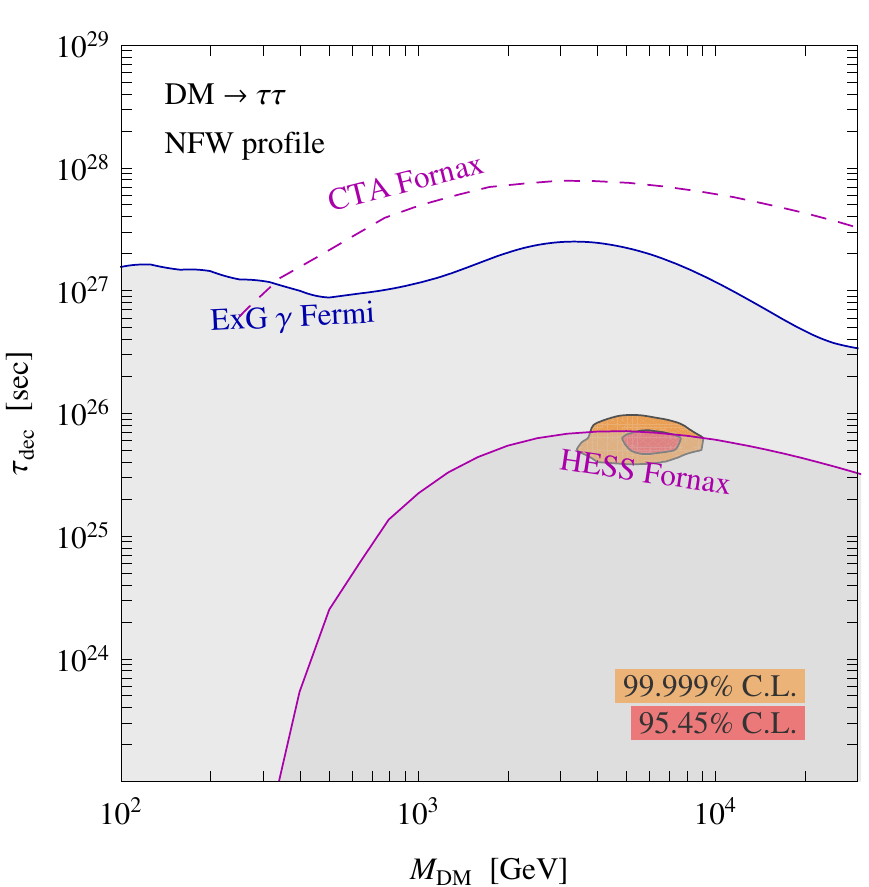}\
\includegraphics[width=0.45\textwidth]{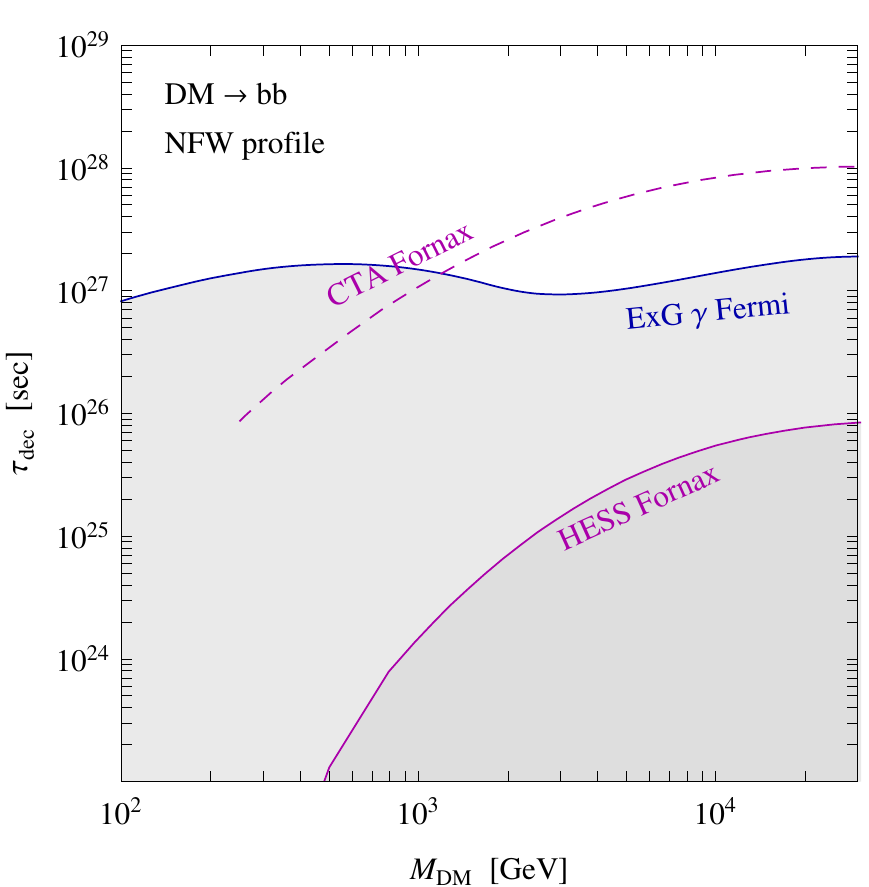}\\[1mm]
\hspace{-8mm}
\includegraphics[width=0.45\textwidth]{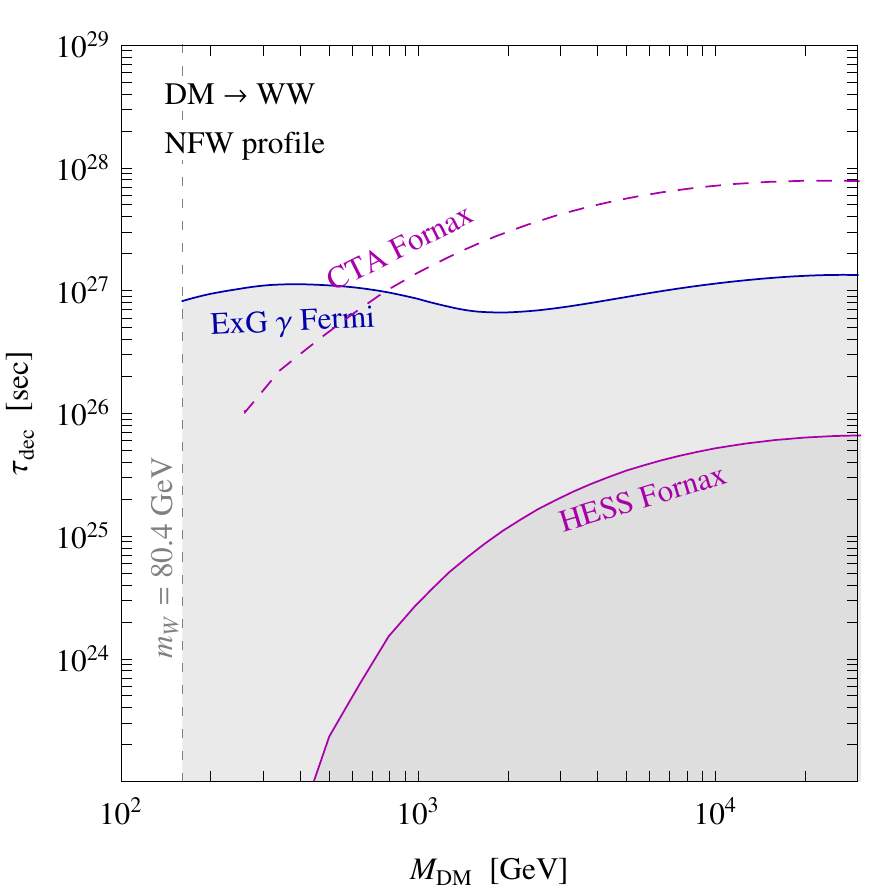}\
\includegraphics[width=0.45\textwidth]{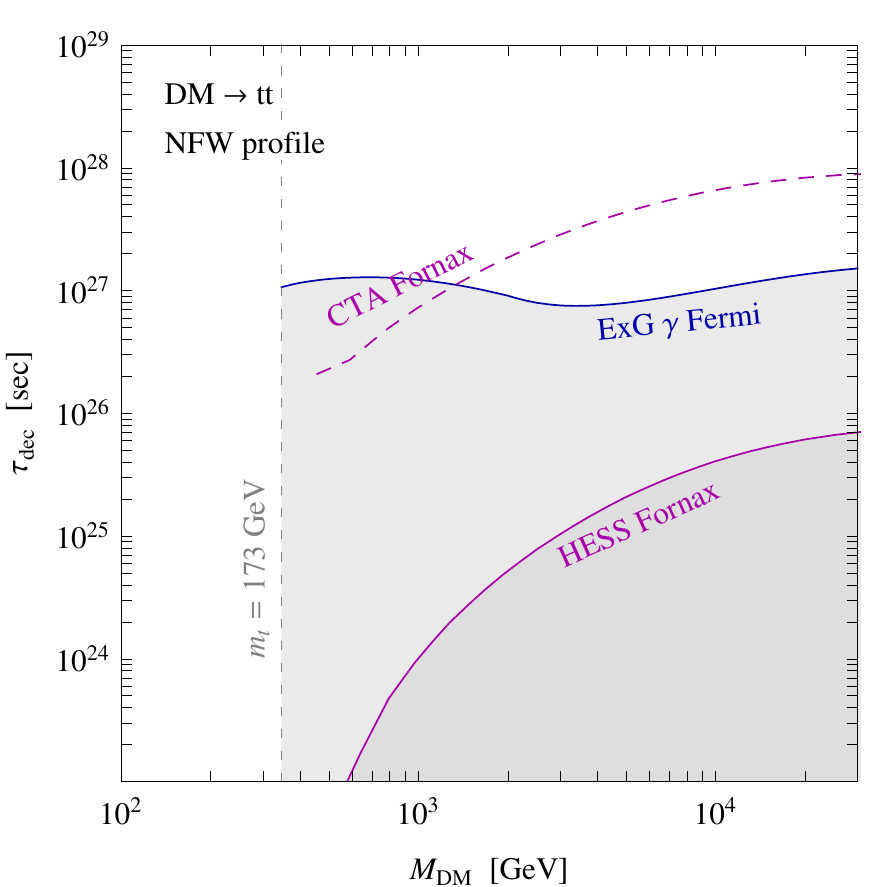}\
\caption{\em\label{fig:exclusion} 
The regions on the parameter space $M_{\rm DM}$--$\tau_{\rm dec}$ that are excluded by the \FERMI\ and \HESS\ constraints and that can be explored by \CTA, together with the regions of the global fit to the charged CR data, for different decay channels.}
\end{center}
\end{figure}

Figure~\ref{fig:exclusion} presents the exclusion plots and therefore summarizes our main results. One can see that the constraints from the \FERMI\ isotropic $\gamma$-ray data exclude decaying DM with a lifetime shorter than $10^{26}$ to few $\times 10^{27}$ seconds, depending on its mass and the precise channel. Therefore, in particular, they rule out the charged CR fit regions, for all the channels. As illustrated in the example in fig.~\ref{fig:exclusionexample}, and inspecting fig.~\ref{fig:example}, adopting the more conservative constraint procedure may marginally reallow a portion of the fit regions, for the DM $\to \mu^+\mu^-$, but leaving a clear tension.
On the other hand, removing the `preliminary' data of~\cite{FERMIexg} and keeping only those published in 2010 in~\cite{Abdo:2010nz} still allows to exclude the CR fit regions (as illustrated in fig.~\ref{fig:exclusionexample} for the $\mu^+\mu^-$; the other channels are less critical and remain safely excluded). 
The constraints from \FERMI\ rise gently as a function of the mass, essentially as a consequence of the fact that the measured flux rapidly decreases with energy. They also depend (mildly, a factor of a few at most) on the decay channel, as a consequence of the different $\gamma$-ray yield and the different shape of the DM signal in each channel.

The constraints from \HESS\ Fornax remain subdominant, roughly one order of magnitude below the \FERMI\ ones. However, for the case of the DM $\to \tau^+\tau^-$ channel, the bound also reaches the CR fit region and essentially confirms the exclusion. The constraints are cut at low masses by the dying out of the sensitivity in \HESS\ to low energy photons. They do not look competitive with respect to \FERMI\ even for larger masses.

\medskip 

In this study we have focussed on 2-body particle-antiparticle decay modes (DM $\to \ell^+\ell^-, q \bar q, W^+W^-$, where $\ell$ is a charged lepton), typical of scalar DM. We do not address 3-body decays such as DM $\to \ell^+ \ell^- \gamma$, since those are model dependent. For fermionic DM, decay channels such as DM $\to \ell^\pm W^\mp$ are possible: these (in first approximation) can be analysed in our framework with a trivial combination of the DM $\to \ell^+\ell^-$ and DM $\to W^+W^-$ channels. Another possibility which has been recently considered is mixed (`democratic') leptonic channels, such as DM $\to$ 33\% $e^+e^-$ + 33\% $\mu^+\mu^-$ + 33\% $\tau^+\tau^-$: these can of course be approximately derived on the basis of our individual bounds.

\subsection{Comparison with existing bounds}
\label{comparison}

We here comment on the relative strength of the constraints in fig.~\ref{fig:exclusion} with respect to bounds from other analyses or other targets.

With respect to the isotropic $\gamma$-ray constraints of~\cite{CPS1}, the bounds derived here are stronger by a factor of 2 to 3. The reasons of this are discussed at length in the text and essentially amount to: updated datasets, more refined DM analyses and the adoption of a more realistic constraint procedure (some of these effects pull towards a weakening while others for a strengthening).

With respect to the work in~\cite{Huang:2011xr}, our constraints from the \FERMI\ isotropic background are somewhat stronger than their corresponding ones (up to about a factor of 5 for the DM $\to \mu^+ \mu^-$ channel), likely due to the several differences in the analysis discussed in Sec.~\ref{introduction}. In addition,~\cite{Huang:2011xr} presents bounds from the observation of the the Fornax cluster by \FERMI: these are less stringent than our \FERMI\ isotropic background constraints but more powerful than our \HESS-based constraints at moderate masses. At the largest masses, our \HESS-based constraints pick up and match theirs (for the $b \bar b$ channel), as expected from the different operational energy range of \FERMI\ versus \HESS.
The bounds from clusters other than Fornax are less powerful, according to the analysis in~\cite{Huang:2011xr}.  Ref.~\cite{Dugger:2010ys} has derived constraints from several clusters using \FERMI\ data too, but their procedure is questioned in~\cite{Huang:2011xr}, and in any case it is now superseded. 
On the other hand, the preliminary constraints shown (for the $b\bar b$ channel only) in~\cite{Zimmer:2011vy}, obtained with a combination of several clusters in \FERMI, exceed our bounds by a factor of 2. 

Bounds from probes other than the isotropic flux and clusters do not generally achieve the same constraining power. E.g. recently~\cite{Ackermann:2012qk} finds less stringent limits for the leptophilic channels, focussing on the Milky Way halo.

\smallskip

Within the context of observations performed by IACT, we note that the decay lifetime constraints obtained with galaxy clusters are stronger than those from dwarf galaxies. Even for the ultra-faint dwark galaxy Segue 1, which is believed to be the most promising dwarf in the northern hemisphere\,\footnote{Ultra-faint dwarf galaxies are particularly interesting due to their high mass-to-light ratio. However the nature of some of them is still under debate due to their similarity of their properties with globular clusters. Due to their low surface brightness and a few tens of member stars,  their dark matter content is still subject to large uncertainties.}, the constraints are 2 orders of magnitude higher for a dark matter particle mass of 1 TeV~\cite{Aliu:2012ga}.

\smallskip

We also estimate that the constraints that we derive are stronger than those that can come from neutrino observation of the Galactic Center (see for instance~\cite{Abbasi:2011eq}). Precisely computing those ones for all channels and in the same analysis framework of this study, however, is beyond the scope of this work, and would also probably benefit from a better knowledge of the \ICECUBE\ data (especially the reconstructed neutrino energy), which is not currently available.

\medskip

In summary: with the possible exception of the preliminary bounds from a combination of clusters by \FERMI\ for the $b \bar b$ channel, the constraints that we derive from the isotropic $\gamma$-ray flux are the most stringent to date.

\subsection{Prospects for improvements}
\label{prospects}

As we have seen, currently the constraints on decaying DM from the \FERMI\ satellite are dominant with respect to those from \HESS. However, while the former may increase its statistics by at most a factor of a few, for the latter there are prospects of developments in the mid-term future.

The next-generation IACT will be a large array composed of a few tens to a hundred telescopes located on two sites, one in each hemisphere~\cite{Consortium:2010bc}. The goal is to improve the overall performances of the present generation: one order-of-magnitude increase in sensitivity and enlarge the accessible energy range both towards the lower and higher energies allowing for an energy threshold down to a few tens of GeV.  From the actual design study of \CTA, the effective area will be increased by a factor 10 and a factor 2 better in the hadron rejection is expected. 


The calculation of \CTA\ sensitivity is performed with a background-only hypothesis, where the non-detection of a signal by \CTA\ would imply that the signal in the ON region consists of only misidentified hadron showers. 
The background level is estimated by integrating Eq. (4) of~\cite{2012ApJ...746...77V} after multiplying it by the \CTA\ effective area. The upper limit on the number of gamma-ray events is calculated at 95\% C.L. according to the method from~\cite{Rolke:2004mj} assuming five OFF regions.
This calculation follows the methodology described in~\cite{2012ApJ...746...77V}, from which the \CTA\ effective area is extracted. 
The 95\% C.L. sensitivity on the decay lifetime is then given by 
\begin{equation}
\Gamma_{\rm dec}^{\rm 95\% C.L.} = \frac{4\pi}{\int_{\Delta\Omega} d\Omega \int_{\rm los} ds \, \rho_{\rm Fornax}[r(s,\psi)]} \times \frac{M_{\rm DM} \,  N_{\gamma}^{\rm 95\% C.L.}}{T_{\rm obs}\, \int_{0}^{M_{\rm DM}/2} A_{\rm CTA}(E_{\gamma}) \, \frac{dN}{d E_{\gamma}}(E_\gamma) \, d E_{\gamma}} \, ,
\end{equation}
where $ N_{\gamma}^{\rm 95\% C.L.}$ is the limit on the number of gamma-ray events, $A_{\rm CTA}$ is the CTA effective area and $T_{\rm obs}$ the observation time.

Figure~\ref{fig:exclusion} shows the 95\% C.L. sensitivity of \CTA\ on the decay lifetime for the RB02 halo profile for 50h observation time and $\rm \Delta\Omega = 2.4 \times 10^{-4}$ sr.

\medskip

To conclude, we also mention that a technique which could allow for significant improvements in the exploration of the parameter space of decaying DM using clusters is the one of stacking the observation of a large number of different clusters, as recently discussed in~\cite{Combet:2012tt}. 
The authors find that improvements of up to 100 can be theoretically achieved, albeit this factor is $\sim$5 for more realistic background-limited instruments.


\section{Conclusions}
\label{conclusions}

Decaying Dark Matter has come to the front stage recently as an explanation, alternative to annihilating DM, for the anomalies in CR cosmic rays in \PAMELA, \FERMI\ and \HESS. But, more generally, decaying DM is a viable possibility that is or can naturally be embedded in many DM models. It is therefore interesting to explore its parameter space in the light of the recent observational results.

We discussed the constraints which originate from the measurement of the isotropic $\gamma$-ray background by \FERMI\ and of the Fornax cluster by \HESS, for a number of decaying channels and over a range of DM masses from 100 GeV to 30 TeV. We improved the analysis over previous work by using more recent data and updated computational tools.

We found (see fig.~\ref{fig:exclusion}) that the constraints by \FERMI\ rule out decaying half-lives of the order of $10^{26}$ to few $10^{27}$ seconds. These therefore exclude the decaying DM interpretation of the charged CR anomalies, (at least) for all 2-body channels, at least adopting our fiducial constraint procedure. The constraints by \HESS\ are generally subdominant. For the DM $\to \tau^+ \tau^-$ channel, they can however also probe the CR fit regions and essentially confirm the exclusion.

With one possible exception for the DM $\to b \bar b$ channel, the constraints that we derive from the isotropic $\gamma$-ray flux are the most stringent to date.

We also discussed the prospects for the future \v Cerenkov telescope CTA, which will be able to probe an even larger portion of the parameter space.


\paragraph{Acknowledgements}
We thank Daniela Borla Tridon for communications on \MAGIC\ data, Carlos de los Heros for communications on \ICECUBE\ data and Alejandro Ibarra for useful discussions. This work is supported by the French National Research Agency ANR under contract ANR 2010 BLANC 041301 and by the EU ITN network UNILHC. PS thanks CERN's Theory Division for hospitality during the development of this work.

\bigskip

\footnotesize
\begin{multicols}{2}
  
\end{multicols}

\end{document}